# Arc statistics with realistic cluster potentials

# I. Method and first results

**Matthias Bartelmann and Achim Weiss**

Max-Planck-Institut für Astrophysik, Postfach 1523, D–85740 Garching, FRG

**Abstract.** We study the capability of a numerically modelled galaxy cluster to produce giant arcs. The cluster is formed from CDM-like density perturbations via gravitational collapse, which is numerically treated with an $N$-body tree code. We also compute the X-ray emissivity of the baryonic intracluster medium under the assumption that this gas is filled into the gravitational potential of the dark cluster material. From the lensing properties of the cluster, from the statistical properties of the giant arcs produced by the cluster, and from comparing the cluster properties derived both from lensing and from the X-ray properties of the intracluster gas, we conclude (1) that clusters may be significantly more efficient for lensing than estimated from their velocity dispersion on the basis of isothermal-sphere models, (2) that the discrepancy between cluster core radii derived from lensing and from their X-ray surface brightness may be understood assuming that the intracluster gas was expelled from the cluster galaxies when the dark 'body' of the cluster had already formed, and (3) that clusters may be efficient in producing giant arcs even if they have extended cores and a fairly shallow surface-density profile. This may mean that arcs may be much easier to produce, and therefore more abundant, than previously estimated.

**Thesaurus codes:** 12.04.1 – 12.07.1

## 1 Introduction

The statistics of giant luminous arcs and arclets [1] is important for our understanding of clusters of galaxies, of the dark matter in the Universe, and of the issue of structure formation. It may provide information about the amount and the distribution of dark matter in clusters (see, e.g., Grossman & Narayan 1989, Hammer & Rigaut 1989, Bergmann et al. 1990, Mellier et al. 1993, Soucail et al. 1993 and references therein), which is related to the question of how much dark matter exists in the Universe and what it consists of. It may offer an answer to the question of how clusters are distributed in redshift (cf. Wu 1993, Bartelmann 1993), which is closely connected to the question of structure formation in the Universe and of the values of the cosmological parameters $\Omega$ and $\Lambda$ (see Richstone et al. 1992, Bartelmann et al. 1993).

---

[1] We adopt here the interpretation of these phenomena in terms of gravitational lensing.

Gravitational lensing is not the only tool to attack these questions. Cluster masses can in principle also be determined through measurements of the galaxy dynamics in clusters or via observations of the temperature of the intracluster medium (henceforth ICM); see, e.g., White (1992) or Sarazin (1992) for recent reviews. Also, the history of cluster formation and the issue of cluster evolution can be studied with much different methods, for instance by determining the evolution of the X-ray luminosity function (Gioia et al. 1990, Henry et al. 1992), the search for cluster substructure (e.g., Geller & Beers 1982, Jones & Forman 1992 and references therein) or theoretically on the basis of specified cosmogonies (e.g., Kaiser 1992 and references therein).

Gravitational lensing is, however, a theoretically simple concept, and does therefore not suffer from many of the uncertainties which seriously hamper most of the methods mentioned above. For mass determinations via galaxy dynamics, for instance, the degree of virialization of the cluster galaxies should be, but is not, known; moreover, the anisotropy of the cluster-galaxy velocity field is important, but difficult to estimate. As highly resolved cluster X-ray maps become available from ROSAT, we are forced to realize that even the Coma cluster, which is probably the best-studied cluster in the sky, and which was for some time considered to be well relaxed, now reveals at least four substructural features in its X-ray surface brightness (Briel et al. 1992).

It would therefore be at least reassuring for different methods aiming at the same goal, if results on cluster mass distributions obtained from giant-arc observations could confirm results achieved otherwise. The situation, however, is mainly just the contrary. For instance, while mass determinations based on arc observations are mostly in rough agreement with those based on the virialization hypothesis, cluster core radii as determined from lensing appear to be systematically a factor of $\simeq 2$ below those determined from X-ray observations (cf. Mushotzky 1993, Tab.2, Hammer 1991, Miralda-Escudé 1993a). This indicates that, although the total cluster mass – at least interior to circles traced by giant arcs – appear to be reasonably well established, the distribution of this mass is disturbingly unclear. It has been argued, and will be confirmed in this paper, that the curvature radius of large arcs provides an upper limit for the cluster core radius. When this is compared to properties of the X-ray emission, one must bear in mind that the total gravitating mass determines the lensing properties of the cluster, while the X-ray emissivity is given by the distribution of the baryonic intracluster medium. These two distributions can differ markedly, and the possible differences may open a way to learn about the history of the baryonic gas.

The statistics of lensing by galaxies, which results in multiple images from single sources, appears to be settled to an astonishing degree. Multiple-QSO surveys (Crampton et al. 1992, Yee et al. 1993, Surdej et al. 1993, Bahcall et al. 1992, Maoz et al. 1992), when interpreted in terms of 'standard' assumptions on the galaxy population in the Universe, are consistent with basically all these assumptions, as has convincingly been shown by Kochanek (1993). Galaxy lenses are modelled as isothermal spheres with small or vanishing core sizes, their luminosity distributions are described by the Schechter luminosity function, and their luminosities are related to their dynamical properties by either the Faber-Jackson (Faber & Jackson 1976) or the Tully-Fisher (Tully & Fisher 1977) relations, and maximum-likelihood estimates of the parameters entering the relations which determine the prescription come out to be in excellent agreement with determinations based on different methods. The same is true for the cosmological parameter $\Lambda$ and for the luminosity function of the sources.

The interpretation of the statistics of giant arcs is far from being comparably successful. Thorough theoretical studies of the statistics of giant arcs (see, e.g., Kochanek 1990, Miralda-Escudé 1991, Miralda-Escudé 1993a,b, Wu & Hammer 1993, and references therein) have arrived at the main conclusions

1. that the observed 'thinness' of arcs requires surface-density profiles of the clusters steeper than isothermal,
2. that the location of arcs in clusters and their curvature radii require small cluster cores, and
3. that the expected number of giant arcs in the whole sky is smaller than the number of arcs known by now, unless either cluster density profiles are steeper than isothermal or cluster core radii are very small.

All these items are consistent with each other, but basically in disagreement with X-ray observations of clusters. Item (3.) becomes even more pronounced when one takes into account the recent announcement of LeFèvre et al. (1993), that a large fraction of X-ray bright clusters contains giant arcs; i.e., the total number of arcs in the sky may be grossly underestimated. This would indicate that also such clusters which are usually considered unable to produce arcs may be potential lenses. If, furthermore, clusters evolve rapidly with redshift (as discussed by, e.g., Gioia et al. 1990, Henry et al. 1992, Wu 1993, Richstone et al. 1992, Bartelmann et al. 1993, Bartelmann 1993), the number of sufficiently compact clusters capable of lensing is even smaller than usually assumed; then, the difficulty in understanding the number of giant arcs is even enhanced.

Three main conclusions can be drawn from this list: First, as mentioned above, determinations of cluster mass distributions from giant arcs are in conflict with the interpretation of X-ray observations; second, we do not know the distribution of galaxy clusters in their parameter space and in the Universe, nor do we know whether the parameters which are regularly used to describe the lensing properties of clusters are really appropriate, and, third, until recently no arc surveys were performed in a statistically well-defined sample of clusters. The arc survey within EMSS clusters (LeFèvre et al. 1993) has only increased the aforementioned difficulties, in that it reports a very high success rate of finding arcs in X-ray luminous clusters, and that the fraction of arcs with large length-to-width ratio is much higher than expected from 'standard' cluster lens models. Hence, it is mainly unclear what might probably be wrong with present models of the gravitationally lensing fraction of the cluster population in the Universe.

All studies of the giant-arc statistics published up to now are based on analytical models for the cluster surface density. Besides that these models may by themselves be inappropriate, it is not straightforward to unambiguously compare their lensing properties with their X-ray emission. Also, if substructured clusters are frequent, as is indicated by recent studies, then we may expect qualitatively different lensing properties than in the relatively simple models which are analytically tractable. Finally, if a large fraction of the cluster population contains substructure, they are asymmetric; in this case, we expect different lensing properties for different projections of these clusters along the line-of-sight. Such projection effects have been suspected to be important, but have not yet been quantified. To avoid misunderstandings, we want to stress that the previous studies reflect the 'state of the art' concerning analytical models of the lensing cluster population. However, these studies have made remarkably clear that 'something is wrong' with arc statistics, and that the problems with these analytical models call for more complicated, and hopefully more realistic, studies of the subject.

To our opinion, the next theoretical step beyond analytical cluster models has to employ galaxy clusters numerically evolved on the basis of assumptions about the cosmogony of our Universe. This work is intended to be a first attempt to proceed in this direction. Here, we use a single cluster evolved within the CDM cosmogony. This construction is briefly sketched in Sect.2.1. From the spatial distribution of massive particles, which results from this simulation, we derive the cluster density distribution and rotate the cluster into a coordinate frame which is appropriate to study projection effects; this procedure is sketched in Sect.2.2. In Sect.2.3, we derive the basic lensing properties of the resulting surface-density distribution. Sect.3 is devoted to the X-ray emitting ICM and its X-ray properties. In Sect.4, we summarize and discuss the cluster properties, before we, in Sect.5, enter into an analysis of the properties of arcs produced by the cluster model. In Sect.6, we summarize and discuss the results.

We consider this study as a first step. For this one cluster model we develop and demonstrate our methods, which will be applied to a sample of clusters in a forthcoming paper. However, the properties of the present cluster grant some insight into systematic effects, and will therefore be extensively discussed. In the forthcoming study, statistical issues will be the key target.

Throughout, we adopt a Hubble constant of $H_0 = 50$ km/s/Mpc, or $h = 0.5$. Where not otherwise stated, the term 'arc' means 'giant arc'; the latter term will be defined in Sect.4. Arclets, although providing another powerful method to derive cluster mass distributions from gravitational lensing (cf. Kaiser 1992, Kaiser & Squires 1993), are beyond the scope of this study.

## 2 Cluster model

### 2.1 N-body simulation

The cluster is numerically modelled starting from CDM initial conditions using the method described in Steinmetz & Müller (1993). Therefore, an initial density field is generated at high redshift ($z \simeq 1000$) with Gaussian random density perturbations drawn from a CDM spectrum (see, e.g., Bardeen et al. 1986) which is normalized on an 8 Mpc/$h$ sphere with biasing parameter $b = 2$. To ensure that the (comoving) simulation volume of $(5/h$ Mpc$)^3$ contains an effective overdensity that can evolve into a cluster, the density field is scaled by an overall factor of 1.1, and the tidal field expected from the mass distribution outside the box is simulated by imposing on the mass distribution in the box an angular momentum such that it rigidly rotates with a spin parameter of 0.08 ; this is consistent with results of larger-sized numerical simulations (e.g. Barnes & Efstathiou 1987). For details on the numerical method, basically an $N$-body tree code, see Steinmetz & Müller (1993); the number of particles is $N = 42211 \simeq 35^3$. The cluster evolution is followed until $z = 0$; however, we restrict this study to the status of the cluster model at $z = 0.4$, where the calculation box has a physical side length of $(5/h)/1.4 = 3.6/h$ Mpc.

### 2.2 Density field, rotation of coordinate frame

The output of the $N$-body simulation is a set of $3N$ particle coordinates $\boldsymbol{x}_i$, $3N$ particle velocity components $\boldsymbol{v}_i$, and $N$ particle masses $m_i$. To construct a density field from this set of discrete particles, we cover the simulation volume $V$ with an equidistant grid of $N_\mathrm{g}$ cells and sum the particle masses into those grid cells where their trajectories end. Since the cell volume $V/N_\mathrm{g}$ is known, we thus obtain the cluster density defined on the prescribed grid. We choose $N_\mathrm{g} = 64^3$, so that the side length of each cell is

$$\left(\frac{V}{N_\mathrm{g}}\right)^{1/3} = \frac{1}{64}\frac{3.6}{h}\,\mathrm{Mpc} = \frac{56}{h}\,\mathrm{kpc}\,, \tag{1}$$

and we shift the coordinate system such that its origin coincides with the center-of-mass of the cluster. $N_\mathrm{g}$ was adapted such that the smallest length scale of the cluster, i.e. its core radius, was stable against changes in the numerical resolution.

This density field is generically asymmetric. For later studies of possible projection effects, we need a well-defined coordinate system. The natural coordinate frame to describe the cluster is given by the eigensystem of its inertial tensor. Therefore, we form the tensor

$$\begin{aligned}\mathcal{M}_{ij} &\equiv \int_{\mathrm{I\!R}^3} d^3x\,\rho(\boldsymbol{x})\left(\boldsymbol{x}^2\delta_{ij} - x_i x_j\right) \\ &\simeq V\sum_{k=1}^{N_\mathrm{g}}\rho_k\left(\boldsymbol{x}_k^2\delta_{ij} - x_k^i x_k^j\right)\end{aligned}\,, \tag{2}$$

where $k$ numbers the grid cells, $\boldsymbol{x}_k$ is the position vector of the center of the $k$-th cell, $x_k^i$ is the $i$-th component of this vector, $\rho_k$ is the density inside the $k$-th cell, and $\delta_{ij}$ is the Kronecker symbol.

Since the inertial tensor is manifestly symmetric, it can be diagonalized. Hence, there exists a coordinate rotation $\mathcal{T}$ such that

$$\mathcal{M}' \equiv \mathcal{T}^{-1}\mathcal{M}\mathcal{T} = \mathrm{diag}(M_1, M_2, M_3)\,, \tag{3}$$

where the $M_l$, $l \in \{1, 2, 3\}$, are the (non-negative) eigenvalues of the inertial tensor, and the transformation matrix $\mathcal{T}$ is determined by the eigenvectors of $\mathcal{M}$. We now transform the spatial coordinates such as to rotate the inertial tensor of the cluster into its eigensystem and enumerate the new spatial directions in ascending order of the corresponding eigenvalues; i.e., direction 3 is along the eigenvector of $\mathcal{M}$ with the largest eigenvalue.

It turns out that the cluster is approximately a prolate ellipsoid, i.e., it is 'cigar-shaped'. Therefore, it appears roughly radially symmetric when viewed along the 3-direction (cf. Fig.2), and elliptical when viewed along the 1- or 2-direction (see Fig.1). In units of the largest, the eigenvalues are $\{0.72, 0.83, 1\}$.

### 2.3 Lensing properties

Having applied the procedure described in the previous subsection, we find three surface-density fields $\tilde{\Sigma}_i(\boldsymbol{x})$ by projecting the density field $\rho$ along directions $i \in \{1, 2, 3\}$ ; i.e., along the principal axes of the cluster's inertial tensor. To avoid discontinuities in the surface densities, which might arise from collecting discrete numbers of particles into grid cells as described before, we smooth $\tilde{\Sigma}_i(\boldsymbol{x})$ by convolution with a Gaussian of width $\sigma$,

$$\Sigma_i(\boldsymbol{x}) = \left[\tilde{\Sigma}_i * G(\sigma, \boldsymbol{x})\right](\boldsymbol{x}), \quad \text{where}$$
$$G(\sigma, \boldsymbol{x}) \equiv \frac{1}{2\pi\sigma^2} \exp\left(-\frac{\boldsymbol{x}^2}{2\sigma^2}\right). \tag{4}$$

We adopt $\sigma = 50$ kpc/$h$ to make the smoothing length comparable to the cell size of the numerical grid.

To obtain convenient dimensionless quantities for the lens equation, we introduce the mass scale

$$\Sigma_{\rm cr} \equiv \left(\frac{4G}{c^2}\frac{D_{\rm d}D_{\rm ds}}{D_{\rm s}}\right)^{-1}$$
$$\equiv \frac{cH_0}{8G} f(z_{\rm d}, z_{\rm s}), \quad \text{with} \tag{5}$$
$$f(z_{\rm d}, z_{\rm s}) \equiv \frac{(1+z_{\rm d})^2 \left(\sqrt{1+z_{\rm s}} - 1\right)}{\left(\sqrt{1+z_{\rm d}} - 1\right)\left(\sqrt{1+z_{\rm s}} - \sqrt{1+z_{\rm d}}\right)};$$

see Schneider et al. (1992) for details on lens theory. The expression for $f(z_{\rm d}, z_{\rm s})$ is valid for an Einstein-de Sitter universe. Introducing numbers, we have

$$\frac{cH_0}{8G} \simeq 5.4 \times 10^{14}\, h\, \frac{M_\odot}{\rm Mpc^2}; \tag{6}$$

i.e., the critical surface density is of the order of one cluster mass per square Mpc. For the redshifts given from the cluster model, $z_{\rm d} = 0.4$, and assumed for the sources to be introduced later, $z_{\rm s} = 2$, we further obtain

$$f(z_{\rm d}, z_{\rm s}) \simeq 14.3, \tag{7}$$

meaning that the actual critical surface density is about a factor of ten higher than given in Eq.(6).

We further introduce an arbitrary length scale $\xi_0$ in the lens plane, which also defines the natural length scale $\eta_0$ in the source plane,

$$\eta_0 \equiv \frac{D_{\rm s}}{D_{\rm d}}\xi_0. \tag{8}$$

Given these mass and length scales, the lens equation reduces to

$$\boldsymbol{y} = \boldsymbol{x} - \boldsymbol{\alpha}(\boldsymbol{x}), \quad \text{where}$$
$$\boldsymbol{y} \equiv \frac{\boldsymbol{\eta}}{\eta_0}, \quad \boldsymbol{x} \equiv \frac{\boldsymbol{\xi}}{\xi_0}. \tag{9}$$

The scaled deflection angle $\boldsymbol{\alpha}(\boldsymbol{x})$ is determined by the integral

$$\boldsymbol{\alpha}(\boldsymbol{x}) = \int_{\mathbb{R}^2} d^2x'\, \kappa(\boldsymbol{x}')\frac{\boldsymbol{x} - \boldsymbol{x}'}{|\boldsymbol{x} - \boldsymbol{x}'|^2} \tag{10}$$

with the convergence

$$\kappa(\boldsymbol{x}) \equiv \frac{\Sigma(\xi_0 \boldsymbol{x})}{\Sigma_{\rm cr}}. \tag{11}$$

We refine the grid on which $\kappa$ is defined by bilinear interpolation to $2048^2$ cells (pixels) to ensure sufficient resolution on the lens plane for later determination of image properties.

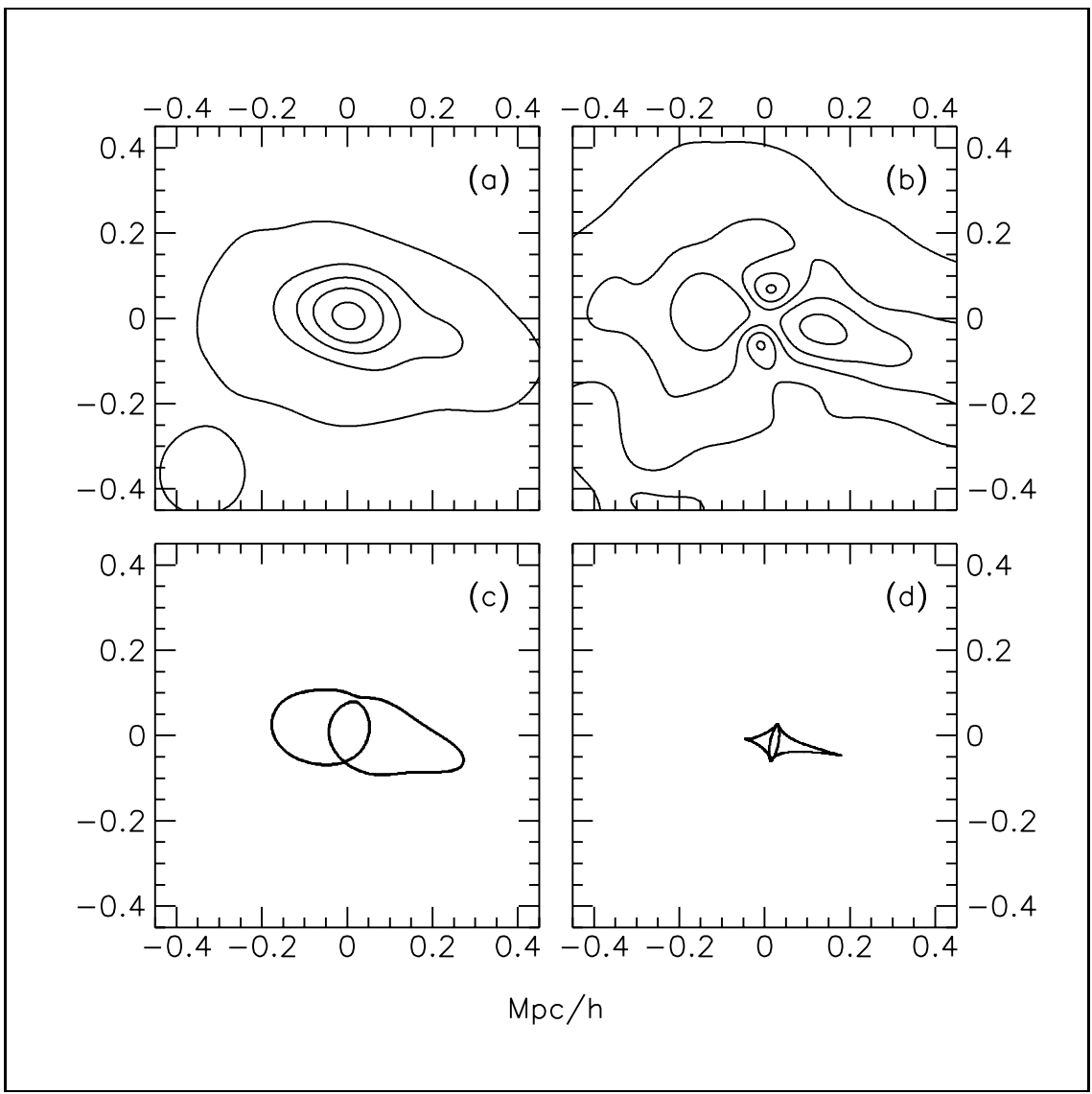

**Fig. 1.** Convergence $\kappa$ (panel a) and shear $\gamma$ (panel b) of the model cluster, projected along the 2-direction, i.e., along that eigendirection of the cluster's inertial tensor with the second smallest eigenvalue. A source redshift of $z_s = 2$ was assumed. Panels (c) and (d) show the critical curve and the caustic of the cluster as seen in the specified direction. The contours in panels (a) and (b) are at $\{10, 33, 50, 67, 90\}\%$ of the maximum, which is $\kappa_{max} = 1.67$ for panel (a) and $\gamma_{max} = 0.61$ for panel (b)

Once the deflection angle is given, the lens mapping is completely determined. It is apparent from Eq.(10) that $\boldsymbol{\alpha}(\boldsymbol{x})$ is a convolution of the convergence $\kappa(\boldsymbol{x})$ with a kernel function

$$\boldsymbol{K}(\boldsymbol{x}) \equiv \frac{\boldsymbol{x}}{|\boldsymbol{x}|^2} , \qquad (12)$$

hence, in symbolic notation,

$$\boldsymbol{\alpha}(\boldsymbol{x}) = (\kappa * \boldsymbol{K})(\boldsymbol{x}) \ . \tag{13}$$

This property of the deflection angle allows to apply fast-Fourier methods to obtain $\boldsymbol{\alpha}$, if $\kappa$ varies slowly over the smallest wavelength introduced into the Fourier transformation; there exists necessarily such a smallest wavelength because $\kappa$ is defined on a grid. Since this is fulfilled in our case because $\kappa$ varies significantly only over several hundred pixels, the convolution theorem applies; i.e.,

$$\hat{\boldsymbol{\alpha}}(\boldsymbol{k}) = 2\pi \ \hat{\kappa}(\boldsymbol{k}) \hat{\boldsymbol{K}}(\boldsymbol{k}) \ , \tag{14}$$

where a hat denotes the Fourier transform and $\boldsymbol{k}$ is the conjugate variable to $\boldsymbol{x}$. The convolution theorem provides the fastest way to compute the deflection angle of a smooth but otherwise arbitrary density field.

From $\boldsymbol{\alpha}$, we obtain the Jacobian matrix of the lens mapping,

$$\mathcal{A}(\boldsymbol{x}) = \left(\frac{\partial \boldsymbol{y}}{\partial \boldsymbol{x}}\right) = \mathcal{I} - \frac{\partial \boldsymbol{\alpha}(\boldsymbol{x})}{\partial \boldsymbol{x}} \ , \tag{15}$$

where $\mathcal{I}$ is the unity matrix. Separating from $\mathcal{A}$ its trace-free part,

$$\Gamma \equiv \mathcal{A} - \frac{1}{2}\mathrm{tr}(\mathcal{A}) \ \mathcal{I} \ , \tag{16}$$

we obtain the shear matrix $\Gamma$ with the shear components

$$\begin{aligned} \Gamma_{11} &= -\Gamma_{22} = -\gamma_1 \equiv \mathcal{A}_{11} - \frac{1}{2}\mathrm{tr}(\mathcal{A}) \ , \\ \Gamma_{12} &= \Gamma_{21} = -\gamma_2 \equiv \mathcal{A}_{12} \end{aligned} \tag{17}$$

and the shear

$$\gamma = -\sqrt{\det(\Gamma)} = \sqrt{\gamma_1^2 + \gamma_2^2} \ . \tag{18}$$

Finally, the magnification factor is determined by the Jacobian determinant,

$$\mu(\boldsymbol{x}) = \frac{1}{\det(\mathcal{A})(\boldsymbol{x})} \ . \tag{19}$$

As an example of the results of this procedure, when applied to our cluster model, we display in Fig.1 the convergence $\kappa_2$ (upper left panel) and the shear $\gamma_2$ (upper right panel), where the subscript '2' now refers to the 2-direction of the projection of the cluster. Additionally, Fig.1 contains the critical curve (lower left panel) and the caustic (lower right panel) determined by convergence and shear.

Fig.1 shows that the cluster is moderately elliptical when seen 'from the side'; the ellipticity of the convergence pattern is

$$\epsilon \equiv 1 - \frac{\text{minor axis}}{\text{major axis}} \simeq 0.3 \tag{20}$$

for the inner contours. The shear reflects this ellipticity: it is highest along the major axis and smallest orthogonal to it. Note in particular the two regions of low shear above and below the cluster center. The shear gives rise to the formation of 'naked cusps', i.e. cusps which lie outside the region of multiple imaging in the source plane. This means that cusp arcs can be formed without causing counter arcs when the source is located

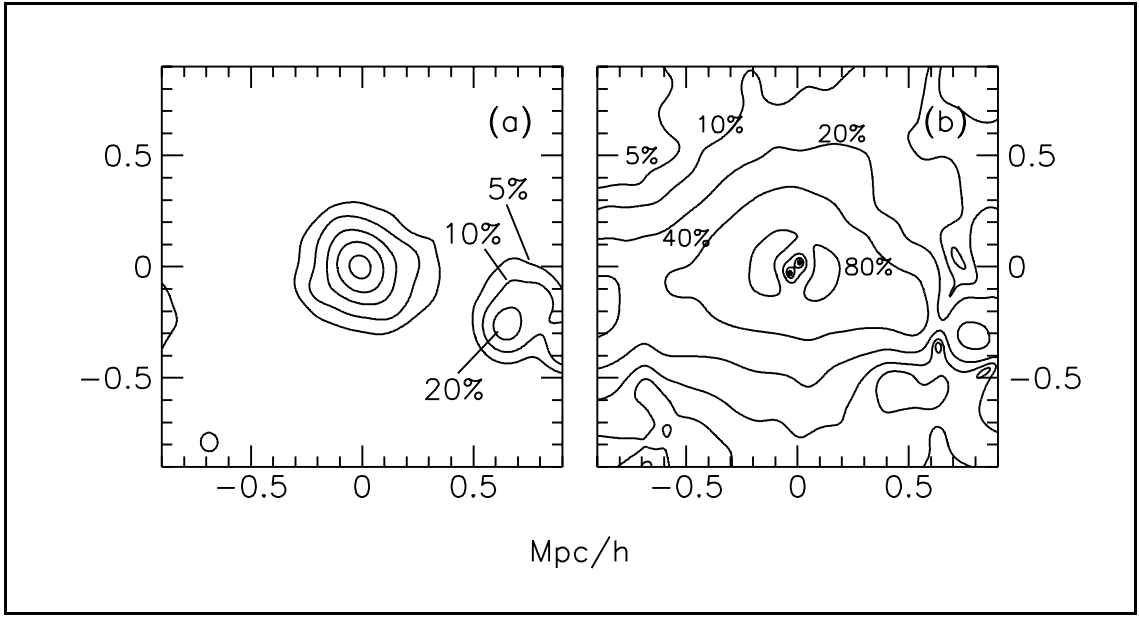

**Fig. 2.** Convergence $\kappa$ (panel a) and shear $\gamma$ (panel b) of the model cluster, projected along the 3-direction; $z_s = 2$. Note that the scale is twice the scale of the contour plots in Fig.1, and that the contour levels are chosen such as to emphasize regions of low $\kappa$ and $\gamma$; the contours are at $\{5, 10, 20, 40, 80\}$ % of the maximum values, which are $\kappa_{\max} = 2.57$ and $\gamma_{\max} = 0.73$. Panel (a) shows that there is a sub-clump of matter to the right of the main part of the cluster, which contains about 10% of the total cluster mass; this clump does not show up when the cluster is seen along direction 2 because it is then projected onto the central part of the cluster. In panel (b), the two features close to the center are minima

close to the naked cusps. The small shear along the minor axis of $\kappa$ causes the tangential critical curve to come close to the cluster center where $\kappa \simeq 1$.

When projected along its symmetric direction (see Fig.2), the cluster appears only mildly elliptical, $\epsilon \simeq 0.1$. There is still a preferred direction in the shear pattern caused by the sub-clump to the right of the cluster center, but this sub-clump affects the critical curves only weakly. When we artificially remove the sub-clump, the radial critical curve is not at all changed, while the tangential critical curve is weakly extended along the direction given by the center of the cluster and the sub-clump. The caustic corresponding to the tangential critical curve is a 'standard', diamond-shaped figure, reflecting the small ellipticity of the convergence. Note that $\kappa_{\max}$ varies by 50% between directions 1 and 2, but $\gamma_{\max}$ only by 10%; the latter is due to the larger ellipticity of the cluster when seen 'from the side'.

## 3 Intracluster gas, X-ray emission

The cluster simulated as described in Sect.2 consists of dark matter; baryonic gas was not included in the simulation. This has a practical and a conceptual reason; the practical reason is the consumption of CPU time, the conceptual reason is that the history of the intracluster gas is not at all clear. From the relatively high iron abundance in the ICM of several clusters (see, e.g., Edge & Stewart 1991), we can conclude that a significant fraction of the ICM was processed by stars, i.e., inside galaxies. At least part of the ICM was therefore expelled by – or stripped from – the cluster galaxies. This makes it less probable that the ICM formed one gravitational potential well together with the dark matter, but rather that at least part of the ICM was 'filled' later into an existing potential well created by the dark matter. The physical processes of how the intracluster gas might have been transported from inside the galaxies into the cluster is unclear. Simulations would have to incorporate reasonably well-resolved model galaxies in a cluster surroundings; a task which is clearly beyond the scope of this work, in particular because we intend to perform statistical studies.

We will therefore assume that the potential well of the dark matter in the cluster can be considered as given, and that the baryonic gas can be treated in hydrostatic equilibrium inside the dark-matter gravitational potential. We further assume that the dominant X-ray emission process is thermal free-free radiation rather than line emission.

### 3.1 Emissivity

The frequency-integrated thermal free-free emissivity of a completely ionized hydrogen plasma is given by (e.g., Rybicki & Lightman 1979)

$$j = \int_0^\infty d\nu \, j_\nu = \frac{8e^6 n_e}{3hc^3 m_e} \left( \sum_i Z_i^2 n_i \right) \sqrt{\frac{2\pi kT}{3m_e}} \,, \tag{21}$$

where we have assumed that the Gaunt factor is reasonably approximated by unity. The sum extends over all types of ions in the plasma with charge $eZ_i$ and number density $n_i$; $n_e$ is the electron number density. Let $\rho_b$ be the baryonic gas density; then, neglecting the electron mass compared to the proton mass $m_p$, we have

$$n_e \sum_i Z_i^2 n_i = \frac{7}{8} \left( \frac{\rho_b}{m_p} \right)^2 \tag{22}$$

for a mixture of 75% hydrogen and 25% helium (by weight). With (22), (21) transforms to

$$j(\rho_b, T) = \frac{7e^6 \rho_b^2}{3m_e m_p^2} \sqrt{\frac{2\pi kT}{3m_e}} \,; \tag{23}$$

i.e., the frequency-integrated thermal free-free emissivity is proportional to the square of the baryonic density and to the square root of the gas temperature.

### 3.2 Baryonic gas distribution

As discussed above, we assume the gravitational potential to be given by the density of the dark matter,

$$\Delta \Phi = 4\pi G \rho_d \,. \tag{24}$$

If the baryonic gas is in hydrostatic equilibrium in that potential, we have, from Euler's equation,

$$\frac{\nabla p}{\rho_b} = -\nabla \Phi \,, \tag{25}$$

where $p$ is the gas pressure. Assuming a polytropic stratification of the baryonic gas with adiabatic index $\gamma$,

$$p = p_0 \left( \frac{\rho_b}{\rho_{b,0}} \right)^\gamma \,, \tag{26}$$

where the subscript '0' refers to the central values of $p$ and $\rho_{\mathrm{b}}$, we obtain

$$\frac{\nabla p}{\rho_{\mathrm{b}}} = \frac{\gamma p_0}{(\gamma - 1)\rho_{\mathrm{b},0}} \nabla \left[\left(\frac{\rho_{\mathrm{b}}}{\rho_{\mathrm{b},0}}\right)^{\gamma-1}\right] , \tag{27}$$

and, after insertion into (25),

$$\frac{\gamma p_0}{(\gamma - 1)\rho_{\mathrm{b},0}} \left(\frac{\rho_{\mathrm{b}}}{\rho_{\mathrm{b},0}}\right)^{\gamma-1} = -\Phi + C , \tag{28}$$

where $C$ is an integration constant. Requiring $\Phi = 0$ where $\rho_{\mathrm{b}} = 0$, i.e., at large distance from the cluster center, we have $C = 0$. Solving (28) for $\rho_{\mathrm{b}}$, we obtain

$$\rho_{\mathrm{b}} = \rho_{\mathrm{b},0} \left[-\Phi \frac{(\gamma - 1)\rho_{\mathrm{b},0}}{\gamma p_0}\right]^{1/(\gamma-1)} . \tag{29}$$

Assuming further that the baryonic gas can be described by the ideal-gas equation, which is very well satisfied because the intracluster gas is very hot and extremely thin, we have for the temperature

$$T = \frac{\mu m_{\mathrm{p}}}{k} \frac{p_0}{\rho_{\mathrm{b},0}} \left(\frac{\rho_{\mathrm{b}}}{\rho_{\mathrm{b},0}}\right)^{\gamma-1} , \tag{30}$$

where $\mu = (4/\sqrt{27}) \simeq 0.77$ is the mean molecular weight of the hydrogen-helium mixture mentioned above.

We put $\gamma = 5/3$ in the following, i.e., we assume that the polytropic gas stratification is adiabatic. Then, we obtain

$$\rho_{\mathrm{b}} \propto (-\Phi)^{3/2} \quad , \quad T \propto \rho_{\mathrm{b}}^{2/3} \propto (-\Phi) . \tag{31}$$

The proportionality $T \propto (-\Phi)$ further clarifies the meaning of assuming an adiabatic gas stratification: for a virialized system of particles, we expect

$$\langle v^2 \rangle \propto (-\Phi) \tag{32}$$

from the virial theorem, $v$ being the particle velocity. If these particles are thermalized, we expect

$$T \propto \langle v^2 \rangle \tag{33}$$

from the equipartition theorem. Combining (32) and (33), we obtain the proportionality derived in Eq.(31) with $\gamma = 5/3$.

Since the X-ray emissivity is determined by $j \propto \rho_{\mathrm{b}}^2 \sqrt{T}$, it follows

$$j \propto (-\Phi)^{7/2} . \tag{34}$$

For comparison, for an isothermal gas, we require $\gamma = 1$ from Eq.(30), and obtain from Eqs.(25,26) $\rho_{\mathrm{b}} \propto \exp(-\Phi)$, or

$$j \propto \exp(-2\Phi) , \tag{35}$$

resulting in a much steeper X-ray surface brightness profile.

To complete the description of the X-ray emissivity of the baryonic gas, we have to specify the central density $\rho_{\mathrm{b},0}$ and the central temperature $T_0$. Observations of the

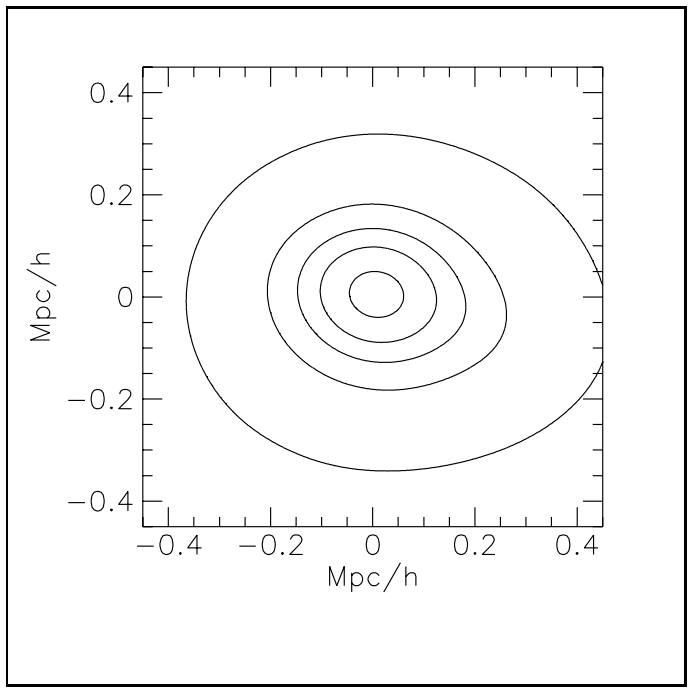

**Fig. 3.** X-ray surface brightness of the model cluster in 2-direction. The contours are at $\{10, 33, 50, 67, 90\}$ % of the maximum, which is $17.4 \times 10^{43}$ erg/s/(Mpc/$h$)$^2$

Coma cluster (White 1992) indicate that the total baryonic mass is $\gtrsim 9$ % of the total cluster mass. Assuming that the baryonic ICM contains roughly half of the total baryonic mass of the cluster, we scale $\rho_{b,0}$ such that the total baryonic ICM mass is 4.5 % of the total cluster mass. If we leave the baryonic fraction of the ICM as a free parameter $\equiv f_b$, the resulting X-ray surface brightness and the total X-ray luminosity would scale $\propto f_b^2$.

To fix the central baryonic gas temperature, we assume that in the center of the cluster the gas is in virial equilibrium with the dark matter, the velocity dispersion of which is known, since we know the particle velocities from the $N$-body simulation. As discussed above, this assumption corresponds to adopting an adiabatic stratification of the ICM. Therefore, we employ

$$\frac{\mu m_p}{2}\langle v^2\rangle_0 = \frac{3}{2}kT_0 \quad \text{or} \quad T_0 = \frac{\mu m_p}{3k}\langle v^2\rangle_0 \,, \tag{36}$$

which specifies the central temperature.

To give an example, we display in Fig.3 the X-ray surface brightness of the cluster, projected along the 2-direction.

### 3.3 Global cluster properties

This subsection summarizes the global properties of the cluster. First, we display in Fig.4 the (azimuthally averaged) convergence and X-ray surface brightness profiles for the three directions of projection.

It is clearly seen from the figure that both the convergence and the X-ray surface brightness are reasonably well described by power laws outside $\simeq 100$ kpc/$h$, while the profiles are almost flat for $\xi \lesssim 30$ kpc/$h$. This means that the cluster shows a pronounced core. The slopes of the profiles can be directly measured; they are contained in Tab.1. Given the slopes, an estimate for the core radii can be found as follows: Let the profile be approximated by the formula

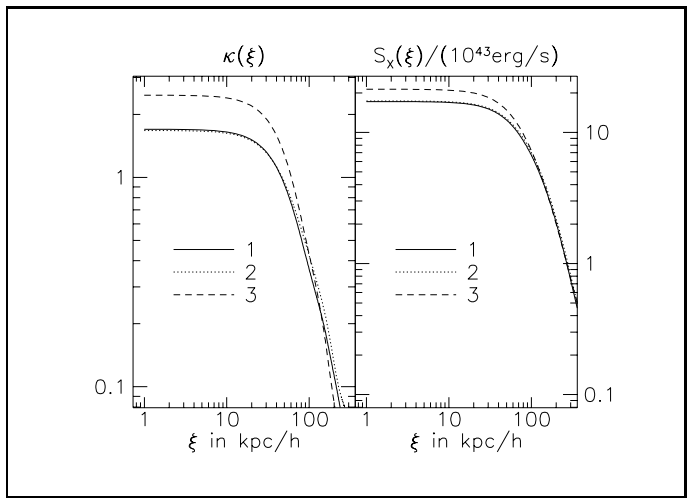

**Fig. 4.** Convergence $\kappa$ and X-ray surface brightness $S_X$ profiles of the model cluster for the three directions of projection (solid line: 1-direction, dotted line: 2-direction, dashed line: 3-direction). The profiles are obtained by azimuthally averaging convergence and X-ray surface brightness

$$f(\xi) = \frac{\xi_c^n f_0}{\sqrt{\xi^2 + \xi_c^2}^n} \;, \qquad (37)$$

i.e., the central value of the profile is $f_0$, its asymptotic slope is $n$, and its core radius is $\xi_c$. Then, at $\xi = \xi_c$,

$$f(\xi_c) = \frac{f_0}{\sqrt{2}^n} \;; \qquad (38)$$

i.e., the core radius is determined by the profile slope and its central value. The core radii determined this way are also listed in Tab.1.

**Table 1.** Properties of the model cluster, dependent on the three directions of projection. A subscript '$\kappa$' refers to the convergence profile, a subscript 'X' to the X-ray surface brightness profile. $n$ is the asymptotic profile slope, $\xi_c$ is the core radius, $\sigma_v$ the velocity dispersion, $S$ the surface brightness, $T_0$ the central temperature, and $L$ the X-ray luminosity

| | direction | | |
|---|---|---|---|
| property | 1 | 2 | 3 |
| $n_\kappa$ | 1.6 | 1.5 | 1.5 |
| $n_X$ | 2.3 | 2.3 | 2.2 |
| $\xi_{c,X}\ h/\text{kpc}$ | 150 | 154 | 133 |
| $\xi_{c,\kappa}\ h/\text{kpc}$ | 81 | 81 | 74 |
| $\sigma_v/(\text{km/s})$ | 736 | 742 | 753 |
| $\kappa_{\max}$ | 1.74 | 1.67 | 2.49 |
| $\gamma_{\max}$ | 0.60 | 0.61 | 0.71 |
| $S_{X,\max}/(10^{43}h^2\text{erg/s/Mpc}^2)$ | 17.3 | 17.4 | 21.4 |
| $T_0/(\text{K})$ | $3 \times 10^7$ | | |
| $L_X/(\text{erg/s})$ | $1.3 \times 10^{44}$ | | |

From $n_X$ and $n_\kappa$, the $\beta$-parameter of the cluster (e.g., Sarazin 1992) can be deduced. By definition, $\beta$ is the squared ratio of the velocities of the galaxies and of the gas particles,

$$\beta \equiv \frac{\mu m_{\mathrm{p}} \sigma_{v,\mathrm{gal}}^2}{kT} \;, \tag{39}$$

where $\sigma_{v,\mathrm{gal}}$ is the velocity dispersion of the cluster galaxies. Solving the spherically symmetric hydrostatic equation for both the galaxies and the ICM, one obtains

$$\rho_{\mathrm{b}} \propto n_{\mathrm{gal}}^{\beta} \;, \tag{40}$$

where $n_{\mathrm{gal}}$ is the number density of the galaxies. If we assume that the (hypothetical) cluster-galaxy distribution would trace the dark-matter distribution, we obtain for $r \gg r_{\mathrm{c}}$,

$$n_{\mathrm{gal}}(r) \propto r^{-2.5} \;, \tag{41}$$

because $n_\kappa \simeq 1.5$, and thus

$$\rho_{\mathrm{b}} \propto r^{-2.5\beta} \;. \tag{42}$$

Since the X-ray emissivity does only weakly depend on the temperature, we approximately have for the X-ray surface brightness

$$S(r) \propto \int dl \, \rho_{\mathrm{b}}^2 \propto r^{-5\beta+1} \;; \tag{43}$$

the integration is taken along the line-of-sight. From Eq.(43), we have

$$n_{\mathrm{X}} = -5\beta + 1 \;, \tag{44}$$

and numerically from Tab.1

$$n_{\mathrm{X}} \simeq -2.3 \;; \tag{45}$$

therefore, we estimate

$$\beta \simeq 0.66 \;. \tag{46}$$

This value of $\beta$ well approximates the value observed on average; Jones & Forman (1984) determine

$$\langle \beta \rangle \simeq \frac{2}{3} \;. \tag{47}$$

For a virialized ICM, one would expect $\beta \simeq 1$. However, Eq.(40), on which the result (46) is based, is valid if the ICM and the galaxies both have an 'isothermal' distribution, i.e., that $T$ and $\sigma_{v,\mathrm{gal}}$ do not vary with $r$, and that the mass is distributed spherically symmetric. The '$\beta$ discrepancy', i.e., the deviation between $\beta = 2/3$ and $\beta = 1$, may therefore indicate that these assumptions are seriously violated in the present case.

The total cluster X-ray luminosity and the velocity dispersion both show that the modelled cluster is rather poor. They are, however, in reasonable agreement with the correlation found by Quintana & Melnick (1982),

$$L_{\mathrm{X}} \simeq 10^{32.71} \, h^{-2} \, \left(\frac{\sigma_{v,\mathrm{gal}}}{\mathrm{km/s}}\right)^{3.94} \frac{\mathrm{erg}}{\mathrm{s}} \;; \tag{48}$$

from that relation, we would deduce an X-ray luminosity of $L_{\mathrm{X}} \simeq 1.1 \times 10^{44}$ erg/s instead of $1.3 \times 10^{44}$ erg/s for the model cluster under the assumption that the galaxy distribution is in equilibrium with the dark-matter gravitational potential.

The cluster core radii, deduced from both the convergence and from the X-ray surface-brightness profiles, show a similar discrepancy like that observed (for a summary, see Mushotzky 1993): the core radius of the dark-matter distribution is lower by a factor of $\simeq 1.9$ than the core radius of the X-ray emitting gas. This statement will be confirmed and discussed below.

# 4 Arc statistics

This section deals with the statistical properties of the arcs produced by the model cluster. We concentrate mainly on three arc properties, namely their width, their length-to-width ratio, and their curvature radius. Where not stated otherwise, we give length and width in units of the assumed source diameter, such that these quantities directly reflect the magnifications along and perpendicular to the arcs. Small source sizes require a highly resolved grid in the lens plane to ensure that the resulting images still contain a sufficiently large number of pixels for the arc recognition algorithm to work reliably. We therefore choose rather large sources, with a diameter of $3''$, to keep the computational costs low, but we have checked in one run of the code with a source diameter of $1.5''$ and twice the standard resolution that the results do not significantly depend on resolution or source size.

To see what happens to the results when the cluster density is artificially enhanced by 25%, we have performed the calculations with both the 'original' and the density-enhanced cluster, referred to in the following by 'model I' and 'model II', respectively.

## 4.1 Arc recognition algorithm

After the lensing properties of the cluster have been determined (Sect.2.3), we are ready to proceed with the next step, which consists of finding the images of a sufficiently high number of sources, to determine their properties and to calculate the probability for long and thin arcs. Miralda-Escudé (1993b, Appendix) already has outlined how to do this. While we follow his prescription in general, our method differs in some aspects mainly because the cluster model is not analytic but is inherently discrete due to its numerical origin.

In Sect.2.3, the deflection angle was determined on a grid of positions $\boldsymbol{x}_i$ ($1 \leq i \leq 2048$) in the lens or image plane. Since the lens mapping is close to the identical mapping along the border of the lens plane because the cluster is concentrated in the center of this plane, we restrict the search for arcs to the central quarter of the grid on which $\kappa$ was defined. Therefore, we use only the $1024^2$-point grid $\boldsymbol{x}_i$ with $513 \leq i \leq 1536$ for arc statistics.

Inserting these positions into the lens equation (9), we obtain the coordinates $\boldsymbol{y}_i(\boldsymbol{x}_i)$ in the source plane onto which these image positions are mapped. We call these coordinates the *mapping table*. From this, we immediately can determine $\det(\mathcal{A})$ by numerical differentiation, and from the sign of this Jacobian determinant we find the critical lines. Naturally, we can give only their approximate location because of the discrete grid. To improve the accuracy, the grid resolution is doubled in such cells which have a nearest-neighboring cell with the opposite sign of the Jacobian determinant. This process is repeated until the critical line (and after mapping the caustics) are found with sufficient accuracy. In that manner, panels c and d of Fig.1 were obtained. In contrast to our approach, Miralda-Escudé (1993b) located the zeroes of $\det(\mathcal{A})$ analytically and followed the lines until a closed curve was found. For our further procedure, the knowledge about the critical lines is of purely diagnostic interest.

To find the images of any given extended source, we employ a method which is based on Schramm & Kayser (1987) and has successfully been used to model the arc in Cl0500−24 (Wambsganss et al. 1989). All image-plane positions $\boldsymbol{x}_i$ are checked whether the corresponding entry in the mapping table $\boldsymbol{y}_i$ lies within the source; i.e., for a circular source of radius $r_\mathrm{S}$ and center at $\boldsymbol{y}_c$, it is checked whether

$$[\boldsymbol{y}_i(\boldsymbol{x}_i) - \boldsymbol{y}_c]^2 \leq r_\mathrm{S}^2 \,. \tag{49}$$

Those lens plane positions fulfilling Eq.(49) are part of one of the source images and will be called *image points* furtheron. The advantage of this method lies in its simplicity and minimal computational effort, but it suffers from the fact that the accuracy of the image positions is always limited by the coarseness of our mapping table. In particular we do not know the positions of the images of the source center.

The next step consists of collecting all those image points that are direct neighbours to each other into one of the 'images', the number of which ranges from 1 to 5 for our cluster. After we have determined the images as lists of image points, it is straightforward to find those points defining the image boundary since they are not surrounded completely by other image points. These boundary points are stored separately for later use.

Miralda-Escudé (1993b) determined the images of any point within a given extended source by solving the lens equation iteratively, making use of his knowledge about the caustics and such about the number of images and their positions. While his method yields the complete information about the images, ours is more practicable for a huge number of sources and is sufficiently accurate for a numerically determined surface mass density. We have also checked the reliability of our method by staring at literally hundreds of arcs and comparing the arc parameters given by this algorithm with our eye's impression.

The area and circumference of an image is immediately given by the numbers of image points (which we associate with a "pixel area") and by the boundary points, along which we just have to integrate numerically. Following again Miralda-Escudé (1993b) we next determine a circle that is defined by three points of the image. These points are (1) an approximation of the source center image, (2) the image point having the largest distance from (1), and (3) the image point having the largest distance from (2). Since we do not know the image of the source center exactly (due to the coarseness of our mapping table), we choose that image point that is mapped closest to it as point (1). For long arcs there might exist three images of the source center and this approach might find a point close to one end of the arc instead of one at the geometrical center. Nevertheless, this does not pose a problem since both points are lying on almost the same circle. The length and curvature radius of the circle segment enclosed by points (2) and (3) are taken as those of the image.

Last, we need an approximation for the image width, which is not uniquely defined. Our approach is to find a geometrical figure with the same area and length as the image. The test for the quality of the geometrical fit is then the agreement between the circumference of the equivalent geometrical figure and that of the image. Our tests yielded that for almost all images an ellipse was appropriate. In this case the image length corresponds to the major and the image width to the minor axis length. Besides the ellipse, rectangle and ring were tested. In $\simeq 0.2$ % of all images, our algorithm could not find a good approximation. These were always cases, when the images were extended and irregularly shaped; e.g., images of sources close to hyperbolic umbilics like that shown in Miralda-Escudé (1993a, Fig.1c). In 84% of all cases the relative deviation between image and figure circumference was less than 5% and in 49% even smaller than 3%.

As a last issue to be considered one has to distribute a large number of sources. Obviously one should distribute less sources in those parts of the source plane that are far away from any caustic, and more sources close or inside the caustics. Our method is similar to that finding the critical lines: we first distribute sources on a coarse, but uniform grid. From our mapping table, we obtain the magnification. If it changes by more than 1 (absolute value) between two sources, we next place an additional source between both, thereby enhancing the resolution by a factor 2 in each dimension. For the n-th iteration of source positions the criterium to add additional ones is that the magnification changes by $2^{n-1}$. In the particular case of the present cluster, 5 iterations of source positions were performed, yielding a total number of $\simeq 2300$ sources for the 3-direction of projection. The smallest distance between source centers was approximately equal to the source radius.

### 4.2 Arc properties

Before going into the details of our statistical results, it is worth recalling some general statements about large arcs.
  1. Large arcs are thin, if $\kappa \lesssim 0.5$ at the position of the critical curve. For a thin large arc, the magnification is not larger than its length-to-width ratio.
  2. The curvature radius of large arcs is approximately equal to the local curvature radius of the critical curve.

The validity of these statements can be seen in the following way:

Let $\lambda_1$ and $\lambda_2$ be the two eigenvalues of the Jacobian of the lens mapping. Then, we have
$$\lambda_1 + \lambda_2 = 2(1 - \kappa) \ . \tag{50}$$
Moreover, since we are concerned with large arcs, one of the eigenvalues is close to zero, because otherwise the arcs would not be large. In particular, if we are close to a critical curve, this is seen from the definition of the latter,
$$\det(\mathcal{A}) = \lambda_1 \lambda_2 = 0 \ . \tag{51}$$
Let this small eigenvalue be $\lambda_1$; then, from (50),
$$\lambda_2 \simeq 2\,(1 - \kappa) \ . \tag{52}$$
Arcs are called 'thin' when they are demagnified in width; therefore, the condition for a thin arc can be written
$$\lambda_2 \gtrsim 1 \quad \text{or} \quad \kappa \lesssim \frac{1}{2} \ , \tag{53}$$
as stated above. Moreover, the magnification of an arc is
$$\mu = \frac{1}{\det(\mathcal{A})} = \frac{1}{\lambda_1 \lambda_2} \quad \text{or} \quad \lambda_1 = \frac{1}{\lambda_2 \mu} \ . \tag{54}$$
The length-to-width ratio of a large arc can be approximated by
$$\frac{L}{W} \simeq \frac{\lambda_2}{\lambda_1} = \mu \lambda_2^2 \simeq 4\mu\,(1-\kappa)^2 \ ,$$
$$\Rightarrow \quad \mu \simeq \frac{(L/W)}{4(1-\kappa)^2} \ ; \tag{55}$$

cf. Wu & Hammer (1993), Eq.(7). For $\kappa \lesssim 0.5$, we therefore obtain

$$\mu \lesssim \frac{L}{W} \,. \tag{56}$$

Hence, the length-to-width ratio of a thin arc is an upper bound to the magnification of the arc. This demonstrates what has been claimed in statement (1.).

As to the second statement, the local tangent to a large arc is approximated by the local direction of the major principal axis of the shear matrix $\Gamma$. Therefore, the angle between the local tangent to the arc and 1-axis of the coordinate system is given by

$$\tan 2\phi = \frac{\gamma_2}{\gamma_1} \,. \tag{57}$$

Upon changing $\gamma_i$ by a small amount, this angle changes by

$$|d\phi| = \frac{|\gamma_1 \, d\gamma_2 - \gamma_2 \, d\gamma_1|}{2\gamma^2} \,. \tag{58}$$

In general, we have

$$d\gamma_i = \nabla \gamma_i \, d\boldsymbol{l} \,, \tag{59}$$

where $d\boldsymbol{l}$ is a small directed path segment. In polar coordinates, this reads

$$d\gamma_i = d\boldsymbol{l} \left[ (\partial_r \gamma_i) \, \boldsymbol{e}_r + \frac{1}{r} (\partial_\phi \gamma_i) \, \boldsymbol{e}_\phi \right] \,, \tag{60}$$

where $(\boldsymbol{e}_r, \boldsymbol{e}_\phi)$ are unit vectors in $(r, \phi)$ direction.

Consider now a small segment of the critical curve and choose the coordinate system such that the 1-axis is normal to the local tangent to the critical curve, and that the coordinate origin coincides with the curvature center of the circle of which the critical curve segment is a part. Since we are interested in large tangential magnification, we have

$$\gamma_2 \simeq 0 \quad \Rightarrow \quad |d\phi| \simeq \frac{|d\gamma_2|}{2|\gamma_1|} \,. \tag{61}$$

If we proceed along the direction tangent to the critical curve, we obtain for $d\gamma_2$

$$d\gamma_2 \simeq (\partial_\phi \gamma_2) \, d\phi \,, \tag{62}$$

or, with (61), we must have

$$\partial_\phi \gamma_2 \simeq 2\gamma_1 \,. \tag{63}$$

Hence, with $d\boldsymbol{l} = dl \, \boldsymbol{e}_\phi$ in Eq.(60), we obtain

$$|d\phi| \simeq \frac{dl}{r} \,. \tag{64}$$

It follows from this equation that the 'curvature radius' determined by the principal axis of the shear matrix is approximately equal to the local curvature radius of the critical curve. Assuming that the curvature of a large arc is determined by the change in direction of the principal shear axis, which should be well satisfied for arcs which are not extremely long, this confirms statement (2.).

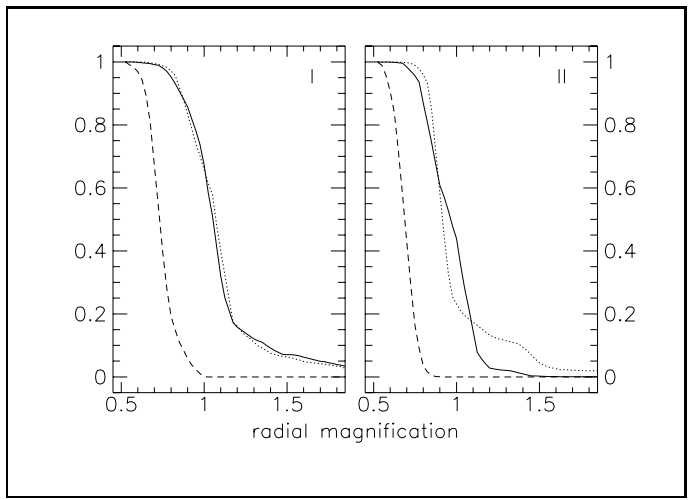

**Fig. 5.** Cumulative probability distribution for the width of large arcs, $L \geq 8$. The left panel shows the results for model I, the right panel those for model II. The three curves in each panel are for the three directions of projection of the cluster; solid line: 1-direction, dotted line: 2-direction, dashed line: 3-direction. The 3-direction is along the eigendirection of the cluster's inertial tensor with the largest eigenvalue, i.e., along the 'most symmetric' direction

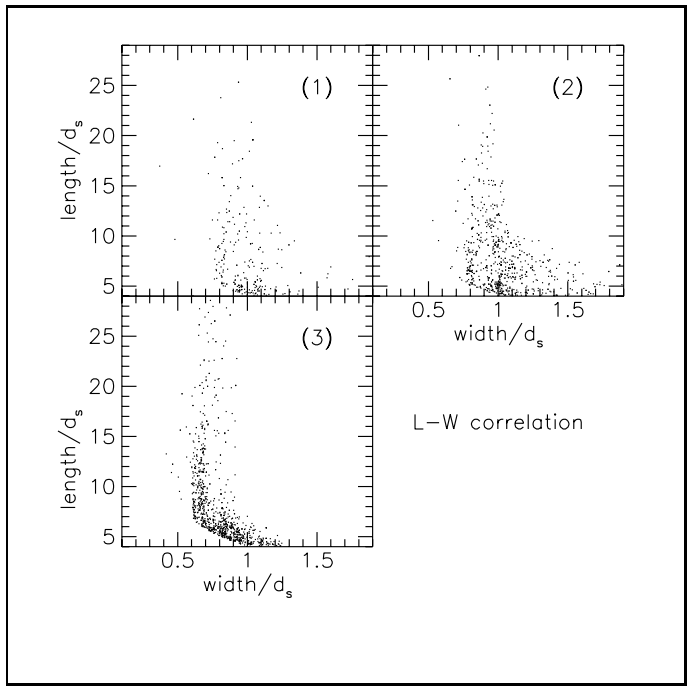

**Fig. 6.** Correlation between length and width for arcs with $L \geq 4$ for model I; the numbers in the panels show the projection direction of the cluster. The plots show that long arcs have $W \simeq 1$ when the cluster is seen 'from the side', while the width decreases from $W \simeq 1$ to $W \simeq 0.6$ for $L \lesssim 7$ and then remains there for longer arcs when the cluster is seen along its 'symmetric' direction. Moreover, the spread in $W$ is larger for directions 1 and 2 than for direction 3

**4.2.1 Occurrence of thin arcs.** Fig.5 displays the cumulative probability distribution of the width of arcs with a length larger than 8.

First, Fig.5 exhibits a significant projection effect. While the median width for the 1- and 2-directions is $\bar{W} \simeq 1.1$ for models 1 and 2, it reduces to $\bar{W} \simeq 0.7$ for the 3-direction; moreover, basically all large arcs are thin in this case while most of them are 'thick' in the other cases. This means that thin and large arcs have a significantly higher probability to be formed when the line-of-sight through the cluster is close to its 'symmetry' axis, where also the central $\kappa$ is largest.

Second, Fig.6 indicates that there is a clear correlation between length and width for the arcs; the width remains roughly constant when the length exceeds $L \gtrsim 10$. This means that basically all the long arcs have roughly the same width, and it also confirms the projection effect seen in Fig.5: The long arcs produced by the cluster, when seen from the side, have $W \simeq 1$ while they are narrower when the cluster is seen along its 'symmetric' direction; then, $W \simeq 0.6$, and the spread in $W$ is also smallest in this case.

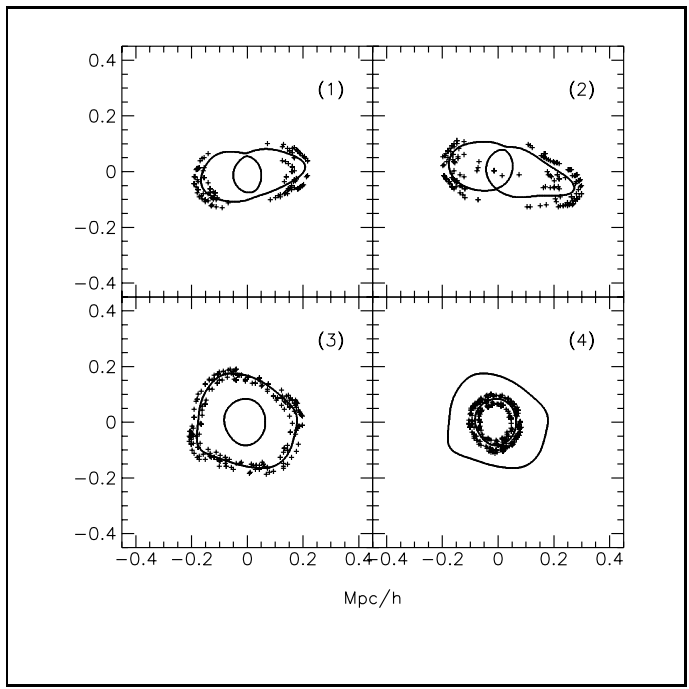

**Fig. 7.** Positions of long ($L \geq 8$) thin ($W \leq 1$) arcs relative to the critical curves for projection along the asymmetric directions (panels 1 and 2) and along the 'symmetric' direction (panel 3). Panel 4 shows the positions of radial arcs, seen along the 'symmetry' axis of the cluster, with $L > 4$. The results were obtained from model I

To further illustrate the properties of the lens mapping in the three projection directions of the cluster, we display in Fig.7 the positions of long ($L \geq 8$) and thin ($W \leq 1$) arcs relative to the critical curves (panels 1 through 3) and the positions of radial arcs, seen along the 3-direction, with $L > 4$ (panel 4).

The figure shows that, with the exception of few arcs in panel 2, that all long and thin arcs are located close to the tangential critical curve of the cluster; i.e., long and thin radial arcs do basically not occur. While these arcs are roughly homogeneously distributed along the tangential critical curve in panel 3 (i.e., when the cluster is projected along its 'symmetric' direction), they occur preferentially along those parts of the tangential critical curve which are the image of the 'naked' cusps of the corresponding caustic for the other two directions of projection; cf. the two lower panels in Fig.1. These 'naked' cusps are a consequence of the considerable ellipticity of the cluster when seen 'from the side'. Moreover, for projections 1 and 2, the tangential critical curve comes close to the cluster core along the direction of the minor axis of the surface-density ellipse. There, $\kappa > 0.5$, which means that thin arcs cannot occur, as shown in Eqs.(50) to (53); see also Fig.7, panels 1 and 2.

**4.2.2 Arc curvature and length.** The distribution of the length-to-width ratio of the arcs, displayed in Fig.8, also reflects the fact that the large arcs are significantly thinner for the projection of the cluster along the 3-direction.

The thinness of the arcs seen in the 3-direction comes from the rather high central value of $\kappa$ in combination with the shear, which is enhanced by the (small) intrinsic ellipticity in combination with a sub-clump of matter aside the main 'body' of the cluster; cf. Fig.2. This issue is further discussed in the final section. When the cluster is seen 'from the side', the shear acts preferentially in the same direction into which the convergence is elongated. Therefore, although this also leads to an expansion of the critical curve, $\kappa$ is still above 0.5 at the critical curve. Moreover, we argue in Sect.5 (see Eqs.(65–67) and Fig.10) that a low central value of $\kappa$, together with an extended core,

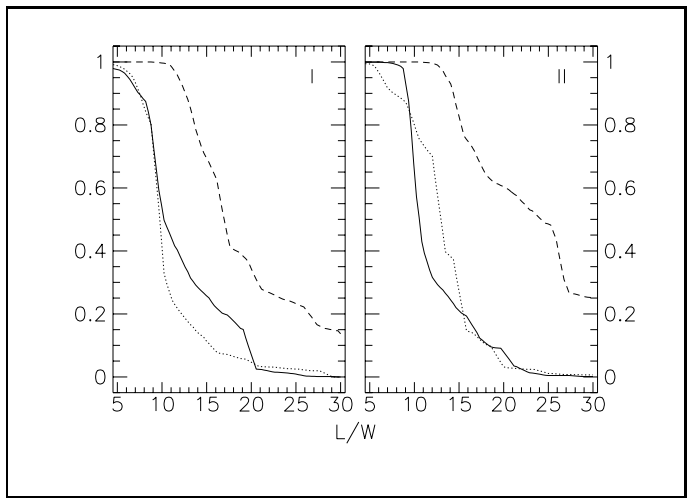

**Fig. 8.** Cumulative probability distributions for the length-to-width ratio of large arcs ($L \geq 8$), produced by the model cluster projected along the 1-, 2-, and 3-directions (solid, dotted, and dashed lines, respectively). The projection effect apparent in Fig.5 also shows up here; large $L/W$ ratios are significantly more frequent among arcs produced by the cluster as seen along the 3-direction. The two panels correspond to models 1 and 2, as indicated.

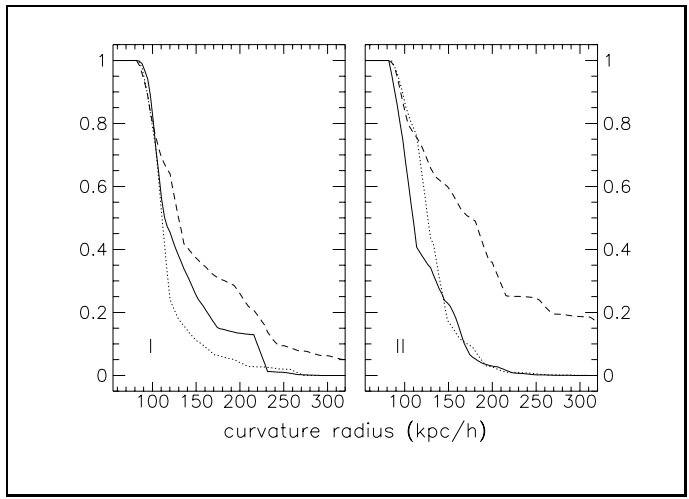

**Fig. 9.** Cumulative probability distributions for the curvature radius of large arcs ($L \geq 8$), produced by the model cluster projected along the 1-, 2-, and 3-directions (solid, dotted, and dashed lines, respectively). For the 3-direction of cluster projection, large curvature radii are more probable than for the 1- and 2-directions, in particular for model 2 (right panel); this is due to the larger fraction of large radial arcs in these cases

make thin arcs less probable.

Figure 2 illustrates the influence of substructures on the lensing properties of the cluster. To the right of the cluster, at a distance of $\simeq 650$ kpc/$h$, there is a clump of matter containing about 10 % of the total cluster mass. This clump, together with the intrinsic shear due to the ellipticity of $\kappa$, enhances the shear of the cluster, thereby extending the critical curves, and causing them to pass through regions where $\kappa \lesssim 0.5$. This, together with the relatively large central value of $\kappa$, makes the large fraction of thin arcs possible although the cluster has a large core and a fairly shallow profile; see also the discussion around Fig.10 in the final section.

As discussed above, the curvature radius of large tangential arcs reflects the local curvature radius of the critical curve at the position of the arc. Since the large arcs are located along strongly curved parts of the tangential critical curve for the 1- and 2-directions of projection (see panels 1 and 2 of Fig.7), their median radius of curvature is smaller than for arcs seen along the 3-direction. This is illustrated in Fig.9.

However, although the fraction of arcs with large curvature radii is larger for the 3-direction, the minimum curvature radius is quite independent of the projection direction. This minimum is at $\simeq 80$ kpc/$h$, which is equal to the core radius of the cluster (see Tab.1). This confirms that the radius of curvature of large arcs is an upper bound to the cluster core radius, as has already been discussed on the basis of analytical cluster models by, e.g., Miralda-Escudé (1993a).

Table 2 summarizes some of the statistical results displayed in the figures of this section.

**Table 2.** Probabilities for the occurrence of long and thin arcs, as well as of arcs with a large length-to-width ratio, for the three projection directions for models 1 and 2

| model | direction | $L > 4$ | $L > 8$ | $L/W > 8$ |
|-------|-----------|---------|---------|-----------|
| 1 | 1 | 0.092 | 0.017 | 0.008 |
| 1 | 2 | 0.085 | 0.023 | 0.009 |
| 1 | 3 | 0.097 | 0.010 | 0.013 |
| 2 | 1 | 0.104 | 0.030 | 0.021 |
| 2 | 2 | 0.128 | 0.032 | 0.026 |
| 2 | 3 | 0.102 | 0.012 | 0.017 |

Note that the probabilities in Tab.2 are related to the set of *all* arcs and not to the set of *large* arcs ($L > 8$) as in Figs.(4,7,8). The table shows that, although thin arcs with a large length-to-width ratio are significantly more frequent among the *large* arcs when the cluster is projected along the 3-direction, the probability for large arcs is smaller by a factor of $\simeq 2$ for the 'symmetric' compared to the 'asymmetric' directions for model 1. Moreover, while the probability for large arcs is significantly increased for the 'asymmetric' directions when the cluster density is increased by 25% (model 2), there is little effect on that probability for the 'symmetric' direction.

## 5 Summary and Discussion

We consider the following list of results to be the main conclusions obtained in this study:
1. The model cluster is approximately a prolate ellipsoid. Its projected ellipticity is $\epsilon \simeq 0.3$ when seen from the side, and $\epsilon \simeq 0.1$ when seen along the 'symmetric' direction. It has a rather low velocity dispersion, $\sigma_v \simeq 750$ km/s, and a rather extended core, $\xi_c \simeq 80$ kpc/$h$. The surface density profile of the cluster is mildly steeper than isothermal, its slope is $n_\kappa \simeq 1.5$.
2. Under the assumptions that
   - the baryonic fraction of the ICM comprises $\simeq 5$% of the total cluster mass,
   - that this gas is in hydrostatic equilibrium with the potential well of the dark cluster material,
   - that it is adiabatically stratified with an adiabatic index of 5/3,
   - that it can be considered an ideal gas, and
   - that the central gas temperature can be estimated by the central virial temperature of the dark cluster particles,
   
   we find that it has a total X-ray luminosity ($L_X \simeq 1.3 \times 10^{44}$ erg/s) compatible with the velocity dispersion derived from Quintana & Melnick's (1982) $\sigma_v$-$L_X$ correlation. The temperature is roughly $3 \times 10^7$ K, the X-ray surface-brightness has a profile slope of $\simeq 2.3$ and a core radius of $\simeq 130$ kpc/$h$. The $\beta$ parameter of the cluster corresponds well to that observed, $\beta \simeq 0.66$ compared to $\simeq (2/3)$.

3. The results from the statistics of large arcs reveal significant projection effects:
   - Thin ($W \geq 1$) arcs are rare among the large ($L \geq 8$) arcs ($\lesssim 20\%$) when the cluster is seen 'from the side', whereas there are no thick arcs when the cluster is projected along its 'symmetry' axis. Also, the median length-to-width ratio of arcs is larger by a factor $\simeq 2$ for the 'symmetric' aspect of the cluster, compared to its 'elliptical' aspects.
   - Considering the curvature radius of the arcs, the projection effect is still present, but less pronounced. There are, however, no arcs with a curvature radius smaller than the (convergence) core radius of the cluster, independent of the projection direction.

It is not at all straightforward to interpret the statistical significance of these results for a sample of clusters. In particular, numbers like the median width, length-to-width ratio, and curvature radius are probably to be taken with care. However, some striking features in the results listed above are probably of relevance for the (statistical) interpretation of arc observations.

First of all, since the X-ray properties of the cluster appear reasonable, we may probably take the X-ray core radius seriously. Then, we reproduce the often-cited discrepancy between the cluster core radius and the X-ray core radius. In our model, this is a natural consequence of our filling the baryonic ICM into the already existing dark-matter potential well. Would we assume instead that the baryonic gas density be proportional to the dark-matter density, as we would expect if the baryonic gas had formed a common potential well with the dark matter, the X-ray core radius would be exactly the same as that deduced from the convergence profile and from arc statistics. Therefore, and because the model for the X-ray gas is the only 'free parameter' of the cluster model, we tentatively conclude that the discrepancy between the X-ray and 'lensing' core radii provides information about the history of the baryonic ICM in that it was not present when the 'body' of the cluster formed, but was filled into the dark potential well later. This hypothesis is also supported by the observation that the ICM contains iron in about one third the solar abundance, which shows that at least a considerable fraction, if not all, of the baryonic ICM was filled into the cluster after being processed in stars, i.e., in the cluster galaxies. Although our assumptions about the properties of the baryonic ICM may at least in part appear somewhat conjectual, the results are not very sensitive upon changes in, e.g., the adiabatic index of the gas. For instance, reducing it from 5/3 to 4/3, i.e., making the stratification of the gas more compact, would reduce the X-ray core radius, but it would still be a factor of $\simeq 1.5$ above the 'lensing' core radius. This change, however, would significantly increase the cluster X-ray luminosity, thereby destroying the compatibility between $L_\mathrm{X}$ and $\sigma_v$. The core radii are insensitive to projection effects, as can be read off from Tab.1.

For a circularly symmetric cluster with a finite core and a convergence-profile slope of $\simeq 1.5$ alike the model cluster, we expect large arcs to be thin only for sufficiently large central values of $\kappa$, despite the slope which is steeper than isothermal. To see this, consider a lens model given by the convergence

$$\kappa(x) = \frac{\kappa_0 x_\mathrm{c}^n}{\sqrt{x^2 + x_\mathrm{c}^2}^n} \; . \tag{65}$$

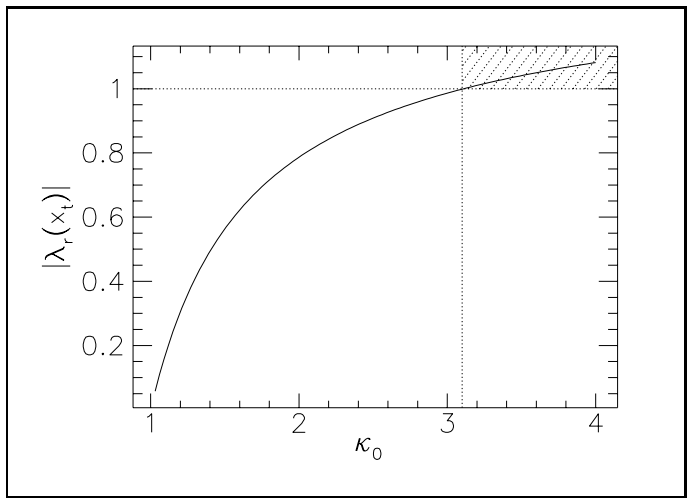

**Fig. 10.** Radial eigenvalue of the Jacobian matrix for the analytic lens model of Eq. (65), $n = 1.5$, as a function of $\kappa_0$; the curve is independent of $x_c$. The shaded area is the region in the parameter plane where thin arcs can be formed. The figure shows that only for $\kappa_0 \gtrsim 3.1$ large arcs are 'thin', i.e., $\lambda_r(x_t) > 1$. Since $\kappa_0 = 2.75$ for the model cluster projected along its 'symmetry' axis, one would not expect thin arcs from this analytical model

Let $m(x)$ be the scaled mass inside a circle of radius $x$,

$$m(x) \equiv 2 \int_0^x dx'\, x' \kappa(x')\;. \tag{66}$$

Then, the eigenvalues of the Jacobian matrix are

$$\begin{aligned}\lambda_r(x) &= 1 - 2\kappa(x) + \frac{m(x)}{x^2}\;, \\ \lambda_t(x) &= 1 - \frac{m(x)}{x^2}\;;\end{aligned} \tag{67}$$

the subscripts 't' and 'r' denote the tangential and the radial eigenvalue, respectively. Fig. 10 displays $|\lambda_r|$ for $n = 1.5$ at the position $x_t$ of the tangential critical curve as a function of $\kappa_0$. The results are independent on the core radius $x_c$, just the tangential critical curve moves outward when $x_c$ increases such that $x_t/x_c = $ const.

It is seen in Fig. 10 that the radial magnification $\mu_r \equiv 1/|\lambda_r|$ at the position of the tangential critical curve is smaller than unity only for $\kappa_0 \gtrsim 3.1$; below, large arcs are not thin despite the fact that the convergence-profile slope is steeper than isothermal. This means that the ellipticity of the cluster, although it is small, makes the thin arcs possible, because a circularly symmetric convergence profile with the same properties as the modelled cluster, in particular with $\kappa_0 = 2.75$, is not able to demagnify large arcs in the radial direction. When the cluster is seen 'from the side', $\kappa_0$ is even smaller, and despite the larger ellipticity the condition for thin arcs is violated along the largest fraction of the tangential critical curve.

The probably most important fact about the model cluster is, however, that its lensing properties are unsuccessfully modelled by an isothermal-sphere lens model. Consider, for example, an isothermal sphere with core, the convergence of which is given by

$$\kappa(x) = \frac{1}{2\sqrt{x^2 + x_c^2}}\;. \tag{68}$$

For such a model, the length scale in the lens plane is

$$\xi_0 = 4\pi \left(\frac{\sigma_v}{c}\right)^2 \frac{1}{f(z_d, z_s)}\;, \tag{69}$$

where $f$ is the redshift-dependent function introduced in Eq.(5). In an Einstein-de Sitter universe with the given redshifts of cluster and source, and with $\sigma_v \simeq 750$ km/s, we obtain

$$\xi_0 \simeq 66 \text{ kpc} ; \tag{70}$$

i.e., the core radius of $\simeq 80$ kpc/$h$ of the model cluster would correspond to

$$x_c \simeq 1.2 . \tag{71}$$

Therefore, a non-singular isothermal sphere with the velocity dispersion and the core radius of the model cluster would be far from being critical; it would reach a central convergence of only $\simeq 0.4$. Conversely, an isothermal sphere with a core radius of 80 kpc/$h$ and a central convergence of 2.75 would have to have a velocity dispersion of $\sigma_v \simeq 2000$ km/s. From the results obtained with the numerically modelled cluster, we therefore tentatively conclude that isothermal-sphere cluster models, which specify the 'lens strength' in terms of a velocity dispersion and a core radius, might grossly underestimate the capability of the cluster population to form arcs. In other words, on the basis of isothermal models, the velocity dispersion of clusters required to produce a significant amount of large arcs may be over-, and the number density of lensing clusters correspondingly underestimated by a large amount.

One might object against this conclusion that the velocity dispersions of some clusters containing large arcs were measured to be large, i.e., $\gtrsim 1000$ km/s. It was however shown by Frenk et al. (1990) that measurements of velocity dispersions in clusters systematically suffer from projection effects and are therefore likely to overestimate the true cluster velocity dispersion.

To conclude, we have to emphasize again that our input cluster model is still somewhat conjectural, in that it was produced from numerical simulations on the basis of the CDM cosmogony, and that we do not know whether this particular cluster's properties are in some sense typical or not. Therefore, the results obtained have to be considered with care. Nevertheless, it seems clear

1. that clusters may be stronger lenses than expected from their velocity dispersion on the basis of isothermal-sphere models, and that clusters containing large arcs may therefore be more abundant than previously thought, therefore providing a possible remedy for the problem with arc abundance emphasized in particular by Wu & Hammer (1993),
2. that the discrepancy between cluster core radii deduced from lensing and from X-ray observations may be understood on the assumption that the major part of the baryonic ICM was expelled from the cluster galaxies when the dark-matter potential well of the cluster had already formed, and
3. clusters can produce thin arcs even if they have extended cores and fairly flat surface-density profiles.

The significance of these results for a statistical sample of clusters will be postponed to a following study.

*Acknowledgements.* We wish to thank Matthias Steinmetz, who provided the numerical cluster model, and Peter Schneider, who triggered numerous enlighting discussions on the subject and who carefully read and commented the manuscript.

# References

Bahcall, J.N., Maoz, D., Doxsey, R., Schneider, D.P., Bahcall, N.A., Lahav, O., Yanny, B., 1992, ApJ 387, 56
Bardeen, J.M., Bond, J.R., Kaiser, N., Szalay, A.S., 1986, ApJ 304 15
Barnes, J., Efstathiou, G., 1987, ApJ 319, 575
Bartelmann, M., 1993, A&A 276, 9
Bartelmann, M., Ehlers, J., Schneider, P., 1993, A&A, in press
Bergmann, A.G., Petrosian, V., Lynds, R.D., 1990, ApJ 350, 23
Briel, U.G., Henry, J.P., Böhringer, H., 1992, A&A 259, L31
Crampton, D., McClure, R.D., Fletcher, J.M., 1992, ApJ 392, 23
Edge, A., Stewart, G., 1991, MNRAS 252, 414
Faber, S.M., Jackson, R.E., 1976, ApJ 204, 668
Frenk, C.S., White, S.D.M., Efstathiou, G., Davis, M., 1990, ApJ 351, 10
Geller, M.J., Beers, T.C., 1982, PASP, 94, 421
Gioia, I.M., Henry, J.P., Maccacaro, J., Morris, S.L., Stocke, J.T., Wolter, A., 1990, ApJ 356, L35
Grossman, S., Narayan, R., 1989, ApJ 344, 637
Hammer, F., Rigaut, F., 1989, A&A 226, 45
Hammer, F., 1991, ApJ 383, 66
LeFèvre, O., Hammer, F., Angonin, M.C., Gioia, I.M., Luppino, G.A., 1993, in: Proceedings 31$^{\text{st}}$ Liège Colloquium on Astrophysics, ed. J. Surdej et al.
Henry J.P., 1992, in: Clusters and Superclusters of Galaxies, NATO-ASI C366, ed. A.C. Fabian (Dordrecht: Kluwer)
Henry, J.P., Gioia, I.M., Maccacaro, J., Morris, S.L., Stocke, J.T., Wolter, A., 1992, MNRAS 386, 408
Jones, C., Forman, W., 1984, ApJ 276, 38
Jones, C., Forman, W., 1992, in: Clusters and Superclusters of Galaxies, NATO-ASI C366, ed. A.C. Fabian (Dordrecht: Kluwer)
Kaiser, N., 1992, ApJ 388, 272
Kaiser, N., 1992, in: Clusters and Superclusters of Galaxies, NATO-ASI C366, ed. A.C. Fabian (Dordrecht: Kluwer)
Kaiser, N., Squires, G., 1993, ApJ 404, 441
Kochanek, C.S., 1990, MNRAS 247, 135
Kochanek, C.S., 1993, ApJ 417, 438
Maoz, D., Bahcall, J.N., Schneider, D.P., Doxsey, R., Bahcall, N.A., Lahav, O., Yanny, B., 1992, ApJ 394, 51
Mellier, Y., Fort, B., Soucail, G., Mathez, G., Cailloux, M., 1991, ApJ 380, 334
Mellier, Y., Fort, B., Kneib, J.-P., 1993, ApJ 407, 33
Miralda-Escudé, J., 1991, ApJ 370, 1
Miralda-Escudé, J., 1993a, ApJ 403, 497
Miralda-Escudé, J., 1993b, ApJ 403, 509
Mushotzky, R., 1993, in: Relativistic Astrophysics and Particle Cosmology, ed. C.W. Akerlof & M.A. Srednicki (New York: Academy of Sciences)
Quintana, H., Melnick, J., 1982, AJ 87 972
Richstone, D., Loeb, A., Turner, E.L., 1992, ApJ 393, 477
Rybicki, G.B., Lightman, A.P., Radiative Processes in Astrophysics (New York: Wiley)
Sarazin, C.S., 1992, in: Clusters and Superclusters of Galaxies, NATO-ASI C366, ed. A.C. Fabian (Dordrecht: Kluwer)
Schneider, P., Ehlers, J., Falco, E.E., 1992, Gravitational Lenses (Heidelberg: Springer)
Schramm, T., Kayser, R., 1987, A&A 174, 361
Soucail, G., et al., 1993, preprint
Steinmetz, M., Müller, E., 1993, in preparation
Surdej, J., Claeskens, J.F., Crampton, D., Filippenko, A.V., Hutsemékers, D., 1993, AJ 105 2064
Tully, R.B., Fisher, J.R., 1977, A&A 54, 661
Wambsganss, J., Giraud, E., Schneider, P., Weiss, A., 1989, ApJ 337, L73
White, S.D.M., 1992, in: Clusters and Superclusters of Galaxies, NATO-ASI C366, ed. A.C. Fabian (Dordrecht: Kluwer)
Wu, X.-P., 1993, A&A 270, L1
Wu, X.-P., Hammer, F., 1993, MNRAS 262, 187
Yee, H.K.C., Filippenko, A.V., Tang, D., 1993, AJ 105 7